\documentclass[sigconf,nonacm,anonymous=false]{acmart}

\usepackage{bm} 
\usepackage{xfrac} 
\usepackage{booktabs} 
\usepackage{algorithm}
\usepackage[noend]{algpseudocode}
\usepackage{caption}
\usepackage[margin=20pt]{subcaption}
\usepackage{hyperref}
\usepackage{enumerate}
\usepackage{mymacro}
\usepackage{balance} 
\usepackage{paralist}
\usepackage{dsfont}
\usepackage{rotating}
\usepackage{listings}
\usepackage{adjustbox}

\usepackage{setspace}

\usepackage{graphicx}
\usepackage[normalem]{ulem}
\useunder{\uline}{\ul}{}

\renewcommand\footnotetextcopyrightpermission[1]{} 

\title{Continuous Attribution of Episodical Outcomes for More Efficient and Targeted Online Measurement}

\begin{document}

\author{Alex Deng}\authornote{Alphabetical order, equal contribution.}
\affiliation{
  \institution{Airbnb}
  \city{Seattle} 
  \state{WA} 
  \country{USA}
  \postcode{98101}
}
\email{alex.deng@airbnb.com}

\author{Michelle Du}
\affiliation{
  \institution{Airbnb}
  \city{San Francisco} 
  \state{CA} 
    \country{USA}
  \postcode{94103}
}
\email{michelle.du@airbnb.com}

\author{Anna Matlin}
\affiliation{
  \institution{Airbnb}
  \city{San Francisco} 
  \state{CA} 
    \country{USA}
  \postcode{94103}
}
\email{anna.matlin@airbnb.com}

\begin{abstract}
Online experimentation platforms collect user feedback at low cost and large scale. Some systems even support real-time or near real-time data processing, and can update metrics and statistics continuously. Many commonly used metrics, such as clicks and page views, can be observed without much delay. However, many important signals can only be observed after several hours or days, with noise adding up over the duration of the episode. When episodical outcomes follow a complex sequence of user-product interactions, it is difficult to understand which interactions lead to the final outcome. There is no obvious \emph{attribution logic} for us to associate a positive or negative outcome back to the actions and choices we made at different times. This attribution logic is critical to unlocking more targeted and efficient measurement at a finer granularity that could eventually lead to the full capability of reinforcement learning. In this paper, we borrow the idea of Causal Surrogacy to model a long-term outcome using leading indicators that are incrementally observed and apply it as the value function to track the progress towards the final outcome and attribute incrementally to various user-product interaction steps. Applying this approach to the guest booking metric at Airbnb resulted in significant variance reductions of 50\% to 85\%, while aligning well with the booking metric itself. Continuous attribution allows us to assign a utility score to each product page-view, and this score can be flexibly further aggregated to a variety of units of interest, such as searches and listings. We provide multiple real-world applications of attribution to illustrate its versatility.  
\end{abstract}

\maketitle

\section{Introduction}

On Airbnb’s marketplace, a booking represents a match between a guest and a listing. Helping our guests find high-quality listings is a continuing effort that is measured with a set of key metrics. Metrics like bookings per user are often either the target metric or one of the guardrail metrics. However, compared to many other user engagement metrics such as product page-views per user and searches per user, bookings per user lacks sensitivity, meaning that it requires more traffic to detect the same effect size. This can be seen from Table~\ref{tab:secompare}, where we compare bookings per user to search and page-view metrics, using the variance of page-views as a baseline. The variance of bookings is 2 to 3 times bigger than searches and page-views, meaning it requires 2 to 3 times more traffic to detect the same effect size as the latter two. A secondary observation is capping counts to 1 and ``binarize'' metric further reduces variance a lot for page-views and searches, but has very limited effect for bookings because most users only book 1 stay for a trip. 

\begin{table}[!htb]
\centering
\resizebox{0.95\columnwidth}{!}{%
\begin{tabular}{@{}lllllll@{}}
\toprule
                        & Booker & Booking & Searchers & Searches & Page-Viewers & Page-views \\ \midrule
Variance of percent lift  & 2.53 &  2.86     & 0.15         & 1.24      & 0.18        & 1        \\ \bottomrule
\end{tabular}%
}
\caption{Comparison of variances from a search ranking experiment, using page-views per user as the baseline.}
\label{tab:secompare}
\end{table}

We see two fundamental challenges for metrics like booking that we want to tackle in this work.
\paragraph{Purchasing decision is episodical and noisy}  First, the metric is based on a final outcome from an often lengthy effort. Unlike page-views, searches and click-through-rates, the booking decision is made only at the very end of the process. Because booking is a big purchase decision (usually at least several hundred dollars), this decision takes time. Moreover, there can be many exogenous factors affecting \emph{whether} the guest will book, \emph{when} the guest will make the decision, and \emph{which} listing the guest will book. For example, a guest may be comparing listings on Airbnb to hotels and other alternatives. Finding a good deal outside of Airbnb could lead the guest to abandon their intention to book a listing on Airbnb. Guests often travel with family or a group of friends, and one guest's booking decision may be influenced by preferences or suggestions that are exogenous to the guest's product experience. Guests also have various degrees of urgency. Those who plan months ahead can gather more candidate listings but delay the booking to a later time. Among guests who eventually booked, it is common that they considered more than one listing, and the motivation for their final selection is not always clear. As the decision process gets longer, more exogenous noises add to the booking outcome. This is in stark contrast to metrics like page-views and click-through-rates, where users' decision and feedback is revealed almost instantly. 

\paragraph{Ambiguous attribution logic} A second challenge is that using booking as the sole signal only attributes reward to a single item/listing. Because conversion rates are typically low for large purchases, this type of attribution results in sparse and noisy data. Many applications such as search ranker offline evaluation with NDCG and online evaluation techniques like interleaving \citep{chapelle2012large} rely on the positive and negative label at the item level. Traditionally, search engines rely on clicks or long-dwelling clicks as a surrogate for relevance. For an episodical conversion scenario such as Airbnb's booking journey, there is no consensus on what can be used as a surrogate for relevance. A natural starting point is to only treat the listing that was ultimately booked as relevant, and to assign all other listings a negative label. A deeper question is: when a guest views a listing multiple times during the search process, should we give more weight to the first impression, or to the views that are closest to the booking decision? These are all real debates, and different choices can lead to differences in measurement efficiency and even opposite evaluation results when comparing two search rankers.

The motivating question for this work is the following: \textbf{Can we design a metric that tracks the causal effect on bookings, with sensitivity closer to page-views and searches, that can also attribute a booking outcome to age-views, searches and other analysis units?}


\section{Methodology}
The central idea is to employ a model-based causal surrogate \citep{athey2019surrogate} to play the role of a value function \citep{sutton2018reinforcement}, and attribute the difference of the value function prior to and post any action at step $t$ as the pseudo-reward received at $t$. 

Let $W_i$ be the binary treatment assignment of subject $i$, and $Y_i$ be outcome at the end of the episode. The causal surrogate model posits the existence of a set of observations $S$ that can block (d-separate\citep{pearl2009causality}) all the causal pathways from the treatment assignment $W$ to the outcome $Y$. This leads to the surrogate assumption of the form of conditional independence in \cite{athey2019surrogate}, 
$$
W_i \indep Y_i | S_i \ ,
$$
which entails that the regression model $\e (Y|S) = \e(Y|S,W=1) = \e(Y|S,W=0)$ applies to both the treatment and control group. This also implies that all the causal effects on $Y$ are mediated exclusively through causal effects on $S$, and pass forward to $Y$ through $\e (Y|S)$. Let $V_i = \e(Y_i|S_i)$, it is straightforward to show that $\Delta(V) = \overline{V}_T-\overline{V}_C$ (T and C means average over the treatment and control group) is an unbiased estimator for the average treatment effect on $Y$. One can also derive the same estimator using the Front-door criteria \cite{pearl2009causality}, and the fact that $W$ is randomization.

$V_i$ is called a surrogate metric or leading indicator metric for a delayed outcome $Y_i$. It is especially critical when $Y$ is a long-term outcome \emph{not observable} for most subjects during the experimentation period. However, in our application, the episodical outcome is delayed but still mostly observed in the experimentation period. The main mathematical property we are exploiting here is the following corollary from the law of total variance:  
\begin{equation}\label{eq:var_red}
\var(V) := \var\{E(Y|S)\} \le \var(Y) \ .
\end{equation}
Because the regression model will potentially smooth out a big portion of $Y$'s variance explained by $S$, we expect significant variance reduction by replacing a highly noisy $Y$ with a regression prediction of it. 

Now we return to the challenge of attributing positive or negative outcomes to individual interactions. Assuming that we identified the surrogate vector $S_t$ at any time step $t$, and computed its causal surrogate $V_t = V(S_t)$, each user interaction with the product updates the surrogate vector to the new state $S_{t+1}$ at $t+1$, with an updated causal surrogate $V_{t+1} = V(S_{t+1}).$ This naturally leads to the attribution of $V_{t+1} - V_t$ ($V_0=0)$) as a pseudo-reward to the user-product interaction making that state update at $t+1$.

\section{More Efficient Conversion Measurement}\label{sec:vr_booking}

To apply this idea to Airbnb bookings, we construct the causal surrogate at each (user, listing) pair level using various user engagements at a central conversion funnel step---the listing page-view. Concretely, we accumulate aggregated actions that a user made on a listing page, factoring in the key components of a guest's exploration experience. Examples of such interactions include viewing photos, amenities, past reviews, and calendar availability. Other notable interactions include contacting the host and clicking the ``Reserve'' button (this click only takes the user to the checkout page with payment details for further confirmation to book, and the conversion rate at the checkout stage varies). We also added listing features such as the listing's location, price, review, availability, past booking history, etc. along with trip planning information such as the number of guests, trip dates and lead time (days to check-in). For training, we collect logs from the preceding 14 days, and for each (user, listing) pair, we label the pair using the observed booking outcome. We train a lightGBM model\citep{ke2017lightgbm} with dropout \citep{vinayak2015dart} for binary classification. 

For scoring, as users are browsing listings in the Airbnb app or website, every new listing page-view marks a new state change and a new time step $t$ of an existing pair, or the beginning of a new pair. We also look back for 14 days to retrieve the cumulative states up to $t-1$ and $t$, with predictions $V_{t-1}$ and $V_t$. The difference $V_t-V_{t-1}$ is attributed to the listing page-view at step $t$. The result is a page-view level attribution, named \emph{listing-view utility}.

To engineer a surrogate metric for booking per user, we sum all the listing-view utilities to the user level, obtaining $U_i$ for user $i$. Unlike bookings per user, which takes integer values and doesn't have much room for truncation or capping, the user-level listing-view utility has a continuous value with a probabilistic scale, representing the propensity of a user to book various listings they have browsed. Since most users only make 1 booking, we cap the user level utility metric at 1 by default. Under further investigation, we found the distribution of user level utility to be highly skewed, and capping it at a lower level such as 0.1 can result in further variance reduction without losing alignment of the signal with the binary booking outcome. Table~\ref{tab:utility_vr} compares the variance of the new utility and capped utility metrics to booking and listing-view metrics. In comparison to the booking metric, the listing-view utility metric reduced the variance by more than \textbf{50\%} (1.24 from 2.86). The capped utility metric reduced the variance further by more than 80\% compared to the booker metric (booking cap at 1), and more than \textbf{85\%} compared to the booking metric.

\begin{table}[!htb]
\centering
\resizebox{0.95\columnwidth}{!}{%
\begin{tabular}{@{}lllllll@{}}
\toprule
                        & Booker & Booking &  Utility Capped &  Utility & Page-Viewers & Page-views \\ \midrule
Variance of percent lift &  2.53 &  2.86     & 0.48        & 1.24      & 0.18        & 1        \\ \bottomrule
\end{tabular}%
}
\caption{The new utility metric reduced variance by more than 50\% compared to booking, and capping further significantly reduced the variance with a combined 85\% reduction!}
\label{tab:utility_vr}
\end{table}

Thanks to Eq \eqref{eq:var_red}, variance reduction doesn't come as much of a surprise. A more important question is whether the causal surrogate assumption holds, and whether the utility metric is truly capturing all or most of the causal effect on booking. We backtested the utility metrics on more than $140$ search ranking experiments from 2021 to 2022 Q1. Naively, one might think that we can just compare or compute the correlation between the two metrics. However, because the booking metric has 2x to 5x more variance than utility, the variance of booking-per-user accounts for the majority of the noise and can easily dilute the correlation. When the effect on bookings is close to 0, the booking metric can even have different signs due to randomness. Therefore, we restricted the comparison only to a subset of 31 experiments where bookings per user have a statistically significant movement indicated by a p-value below 0.05. 

\begin{figure*}[tbh]
    \centering
    \includegraphics[width=0.9\textwidth]{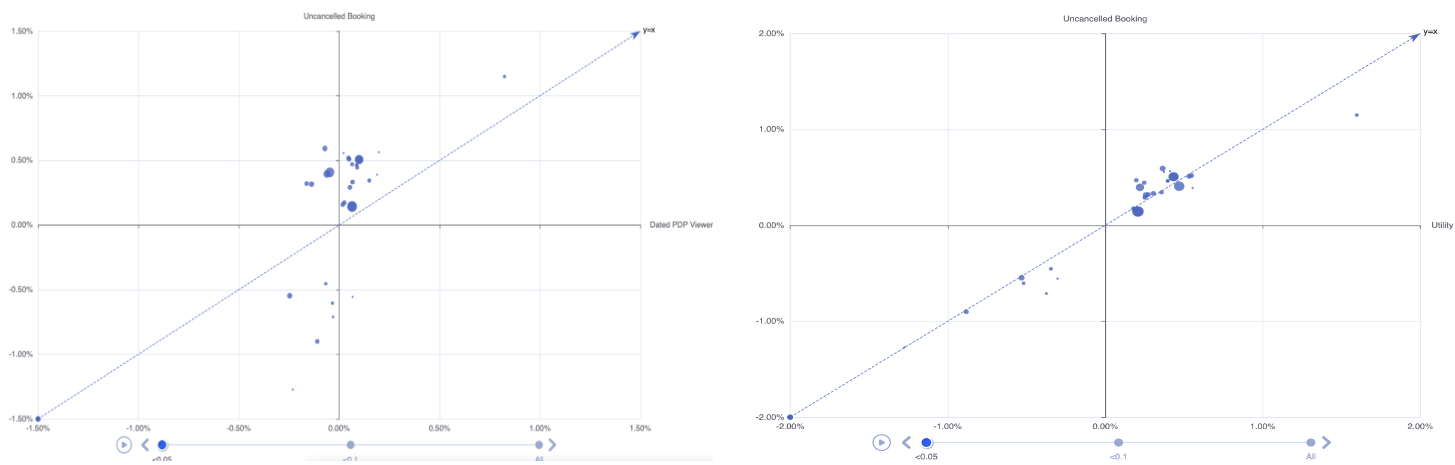}
    \caption{Percent lift of booking (y-axis) vs. listing page-view (left) and utility (right). We only showed 31 out of 140 past search ranking experiments where bookings per user are statistically significant. The size of the dots represents experiment sample sizes.}
    \label{fig:utility_align}
\end{figure*}

Figure~\ref{fig:utility_align} plots dated listing page-viewers (left) and utility (right) on the x-axis against bookings-per-user on the y-axis. Each point represents a single experiment, and the size of the point represents the sample size of the experiment. A diagonal line was added as a reference for perfect agreement. Prior to the utility metric, dated page-viewers (users who viewed a listing with check-in and out dates) was the best surrogate metric from a previous meta-analysis. When comparing to dated listing page-viewers, we found that the utility metric not only improves directional agreement, but also point estimates. The larger the sample size, the closer the booking metric to its ground truth, and utility and booking are also closer. The correlation between the two metrics among these 31 experiments was \textbf{97\%}.

\section{Booking attribution for search level randomization and interleaving}

At Airbnb, online experiments in the search ranking domain are typically implemented with a randomization unit of a logged-in user or a visitor. Experiment traffic is shared among multiple treatment arms, so a given ranking model is typically limited to 25\% of online traffic. In this setting, variance reduction for key metrics translates to faster iteration and a better search experience. 

An exploratory experiment was run with the search as the randomization unit, where the goal is to tune two parameters in a 5 by 5 grid search with a 25\% traffic split into 25 equally sized treatment buckets containing 1\% of all online traffic. It is expected that such a finely granular randomization design will improve statistical power for search level metrics, such as click-through-rate. However, since our main target is booking, a key challenge remains: how do we define a search level metric that can indicate user level impact? The same attribution challenge arises in interleaving\citep{radlinski2013optimized}, where two ranked lists were interleaved into one and we seek to use clicks from user as reward to a winning ranker(in a common team-draft interleaving, the team drafted the item gets the reward). Because the reward is an episodical outcome, it is unclear how to attribute the reward to clicks on the search results. 

Before the introduction of the utility metric, several approaches to attribution were proposed, and they all focused on the booked listing. If a user didn't book in the end, then all engagement signals from their exploration process remained untapped. For the bookers metric, only the booked listing was considered relevant. Because a user often searches and clicks on a listing multiple times before booking, we had a choice to assign the reward to \textbf{all clicks} on the listing result evenly, or to only award the \textbf{first} click. 

By comparison, the listing view utility metric offers a data-driven way to attribute a booking to each search. It also attributes to each search result for interleaving, and can reward the ranker that drafted the listing. Moreover, it distributes reward according to how much information the user gathered on the page toward their booking propensity. Importantly, this also applies to \emph{almost-bookings} -- those listings that the user seriously considered, and compared against, but didn't book in the end. Thus it is more \textbf{efficient} to gather signals from the vast majority of unbooked listings, and at the same time more \textbf{targeted} by assigning reward to searches and search results that lead to the greatest elevation of booking intention. 

\begin{figure*}[bthp!]
    \centering
    \includegraphics[width=\textwidth]{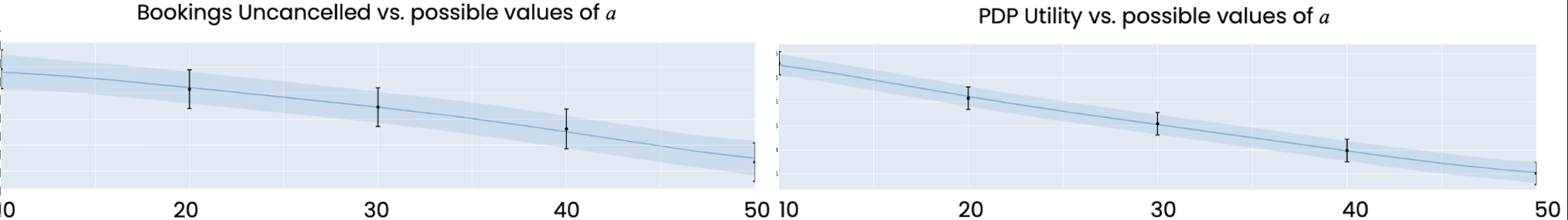}
    \caption{1d parameter tuning of one of the parameter. The utility attributed search level metric (Right) shows a narrower confidence interval comparing to the traditional booked listing clicks per search (left) and can detect the decreasing trend faster. The variance reduction here is 85\%. }
    \label{fig:search_level}
\end{figure*}

Figure~\ref{fig:search_level} displays the point estimate and confidence intervals of booked listing clicks per search (left) and the new listing view utilities per search metric, as functions of varying parameter $\alpha$. We can see that both share the same decreasing trend, but the booked listing clicks per search metric has a much wider confidence interval than its utility counterpart. The variance of the percent change of the new utility based metric is \textbf{85\%} smaller than booked listing clicks per search.

For interleaving experiments, we saw that the utility metrics align with the booking attributed interleaving metric \textbf{78\%} of the time with an additional \textbf{6x} speed up on average. When comparing results in the experiment that also had A/B runs as the ''ground truth'', we see that the alignment is higher at \textbf{89\%} - the few discrepancies observed are likely due to false positives based on our understanding of the experimented features. 89\% alignment between interleaving and A/B tests is considered very high comparing to reported results in \cite{schuth2015predicting}.

\begin{figure*}[thp!]
    \centering
    \includegraphics[width=\textwidth]{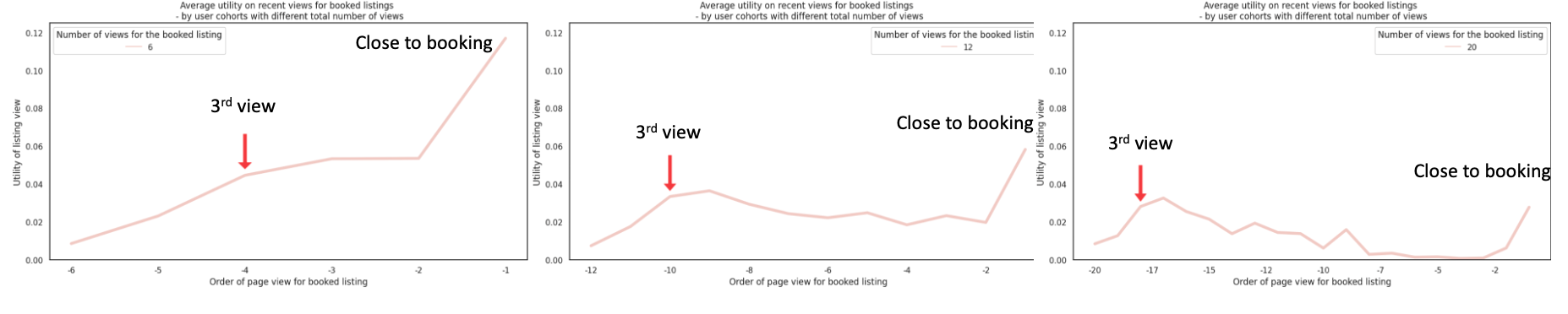}
    \caption{First uptick of utility happens usually around the 3rd and 4th page-views, not the first page-view.}
    \label{fig:user_utility_pickup}
\end{figure*}

\paragraph{Is first impression more important than follow-up views?} To answer this question, we looked at booker cohort by the number of page views on the booked listing. Figure~\ref{fig:user_utility_pickup} shows three cohorts, those with 6 page-views (left), 12 page-views (middle) and those with 20 page-views (right). We plot the 75th percentile of listing view utilities for each page-views from first to the last before booking request. (The utility distribution is highly skewed. A similar trend can be shown with the average or the median, but the 75th percentile manifests the uptick of utility better.) We found that page-views close to booking request have high utility (not surprisingly). In contrast to the common belief that the first impression is very important, the results show that first listing views often do not yield high utilities, and the first uptick of utility happens at 3rd page-view of the same listing. Further looking into the logs reveals that many first listing page views only involve some photo scrolling without more meaningful engagements such as looking at reviews, amenity, long descriptions, etc. Results show two modes in utility attribution: a peak around 3rd and 4th and a second peak close to booking decision. This discovery has direct impact on how we value \textbf{retargeting} --- upranking results that a user has repeatedly clicked will lead to further page-views and may improve conversion up to a point. For the 20 page-view cohort, it is interesting that the utility has a decreasing trend after 5 views, and the booking decision takes place after a gap in activity. This means that the user had already gathered sufficient information for this listing to be a candidate after about 10 page views.

\section{Analyzing user concentration pattern}
Although listing-view utility was designed as a causal surrogate, being able to attribute progress towards a future episodical outcome also introduces analytical opportunities to study purchasing decisions. First, we looked at the concentration of user engagements on the booked listings. The baseline choice is to use the number of listing page-views to represent user intention, and to plot the shares of page views spent on the booked listing among the total page views on each day as it progresses towards the day of booking. Alternatively, instead of raw page-view, we can use the listing view utility to further differentiate the page-views on a continuous scale. Intuitively this kind of shared measures how much a user started to concentrate on the item they want to purchase. 

\begin{figure}[thp!]
    \centering
    \includegraphics[width=0.95\columnwidth]{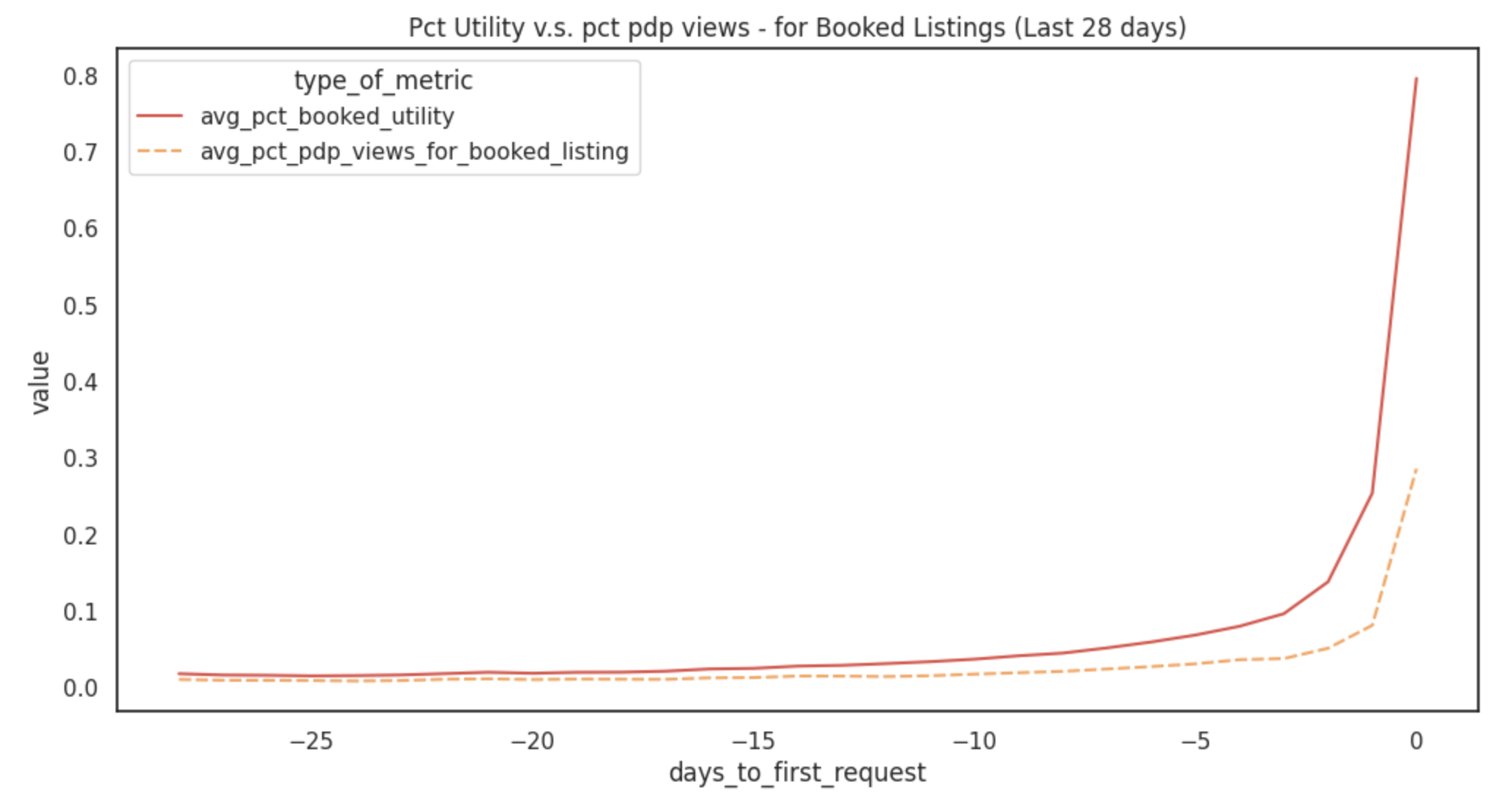}
    \caption{Time trend of the utility share for booked listings (dash) v.s. the page-view share for booked listings (solid) starting from 28 days before the first booking request. Utility share of booked listings show an earlier uptick than the raw page-view share. }
    \label{fig:util_trend}
\end{figure}

Figure~\ref{fig:util_trend} plots the share of booked listing views and the share of utilities at various days to booking, 0 being the day of booking, and -25 being 25 days ahead of booking. \cite{lo2016understanding} used a similar page-view share based metric and found user's purchasing intent start growing significantly from 3 days prior to the purchase. Here we had the same findings using shares of booked listing views. However, when using share of booked listing's utilities we noticed an uptick of the share from 6 days prior to the purchase or even earlier, when the utility share of the booked listing were often 3 times or more than the raw page-view share. 

\section{Related Work}
The theoretical underpinning of our method has been established in \citet{athey2019surrogate} and a special case of front-door criteria in \citet{pearl2009causality}. \citet{athey2019surrogate} also noted the efficiency gain, although their main objective was to study long-term outcomes that are not  observable in a short experiment period (e.g. effect of job training on employment). The focus of our method is to exploit the variance reduction, even when the delayed outcome is short-term and can be observed in a normal experiment period. In particular, we use the model prediction even for subjects with an observed outcome $Y$. We further use the causal surrogate model as a critic model to approximate the value function in the context of reinforcement learning \citep{sutton2018reinforcement} for incremental reward attribution. \citet{drutsa2015future} presented a similar work with the variance reduction angle, but focused on predicting a future value of the same metric that is already continuously observed, e.g. sessions-per-user after 2 weeks using 1 week's outcome. Their work shares the same source of variance reduction as ours and causal surrogacy in general --- due to smoothing effect of conditional expectation, but without the surrogate modeling and cannot be used for episodical outcome. In A/B testing literature, most variance reduction work has been focused on exploiting pre-assignment covariates\citep{deng2013cuped,xie2016improving,lin2013agnostic}.

For delayed outcome and signal, there are also prior works modeling the delay or jointly modeling delay and binary prediction as training with missing data \citep{wang2022adaptive,chapelle2014modeling}. Those works aim at improving prediction of a delayed outcome and evaluated their methods based on accuracy of their forecast. Our work differs in the main objective as not to predict a future metric value but to better estimate the treatment effect, and evaluate our methods based on efficiency gain and empirical alignment with existing goal metrics' effect estimation. Missing data is not a core challenge in our application.  

The idea of using different randomization units in A/B testing and the general rule of thumb of improved efficiency using a finer granular randomization unit whenever possible has been mentioned in earlier works \citep{choiceofexp,Deng:2017}. This directly motivated our desire to seek more granular attribution to allow sub-user level randomization design, and unlock adaptive experimentation opportunities ~\citep{letham2019constrained}.

\section{Conclusion}
In this work we took a model-based causal surrogate to play the role of a value function representing the progress towards a future episodical outcome. By leveraging user's activity on listing page views as surrogate signal, we demonstrated using a surrogate model predicting future conversion successfully led to  50\% to 85\% variance reduction comparing to metrics based on end of episode observed conversions. Using the causal surrogate model as value function, we can attribute incremental gain to each new listing page view, and to all user-product interactions such as searches that leads to listing views. This enables us to experiment and evaluate at a finer granular level below individual user level and still can optimize for a user level episodical outcome. This fine granular attribution also allows us to understand the value of retargetting in a data-driven way, and opens new opportunities to leverage a continuous leading indicator for personalisation.

\balance

\bibliographystyle{ACM-Reference-Format}
\bibliography{ref}

\end{document}